**Superconductivity in REO$_{0.5}$F$_{0.5}$BiS$_2$ with high-entropy-alloy-type blocking layers**

Ryota Sogabe[1], Yosuke Goto[1], Yoshikazu Mizuguchi[1]*

1. Department of Physics, Tokyo Metropolitan University, 1-1, Minami-osawa, Hachioji 192-0397, Japan.

* Corresponding author: Yoshikazu Mizuguchi (mizugu@tmu.ac.jp)

**Abstract**

We synthesized new REO$_{0.5}$F$_{0.5}$BiS$_2$ (RE: rare earth) superconductors with the *high-entropy-alloy-type* (HEA-type) REO blocking layers. According to the RE concentration and the RE ionic radius, the lattice constant of *a* systematically changed in the HEA-type samples. A sharp superconducting transition was observed in the resistivity measurements for all the HEA-type samples, and the transition temperature of the HEA-type samples was higher than that of typical REO$_{0.5}$F$_{0.5}$BiS$_2$. The sharp superconducting transition and the enhanced superconducting properties of the HEA-type samples may indicate the effectiveness of the HEA states of the REO blocking layers in the REO$_{0.5}$F$_{0.5}$BiS$_2$ system.

Since the discovery of the cuprate superconductors and FeAs-based superconductors [1,2], layered compounds have been extensively studied as a potential system in which high-transition-temperature (high-$T_c$) superconductivity and unconventional mechanisms of superconductivity emerge. Basically, layered superconductors have a structure based on alternate stacks of electrically conducting layers, such as the $CuO_2$ layer and FeAs layer, and insulating (blocking) layers, which are typically composed of metal oxides.

In 2012, we discovered new layered superconductors with the $BiS_2$-based conducting layers [3-5]. Basically, the parent phase of the $BiS_2$-based superconductors is a semiconductor with a band gap [6,7]. When electron carriers were generated in the $BiS_2$ layers, superconductivity emerges. In a typical system $LaO_{1-x}F_xBiS_2$, the carrier amount can be controlled by the composition of F ($x$) in the LaO blocking layer [4]. After the discovery in 2012, many $BiS_2$-based superconductors with different types of the blocking layer have been synthesized, and the $T_c$ largely enhanced by replacing (or partially substituting) elements in the blocking layer [5]. Particularly, as reported in Refs. 8 and 9, the relationship between the superconducting properties and the crystal structure have been systematically studied by tuning the chemical pressure effect in the $REO_{0.5}F_{0.5}BiS_2$ system, where the average ionic radius of the RE site was systematically changed by the solution of La, Ce, Pr, Nd, or Sm. Based on those investigations on chemical pressure effects, we found that both carrier doping and the optimization of the local structure (particularly, the suppression of in-plane disorder of the BiS plane) were essential for the emergence of bulk superconductivity in the $BiS_2$-based superconductors [8-12].

The chemical pressure effect in $REO_{0.5}F_{0.5}BiS_2$ is basically controlled by the *alloying effect* at the RE site of the blocking layer. The systematic shrinkage of the blocking layer affects the Bi-S1 (S1 is an in-plane sulfur as depicted in the inset of Fig. 1.) bond distance and induces bulk superconductivity. This fact suggests that the structural properties of the RE(O,F) blocking layer affect the superconducting properties in the $RE(O,F)BiS_2$ system. Furthermore, if the structural stability of the RE(O,F) blocking was enhanced, the superconducting properties may increase in $RE(O,F)BiS_2$. To investigate this possibility, we have studied the crystal structure and the superconducting properties of $(La,Ce,Pr,Nd,Sm)O_{0.5}F_{0.5}BiS_2$ where the $(La,Ce,Pr,Nd,Sm)O_{0.5}F_{0.5}$ layer was designed with the concept of *high entropy alloy* (HEA), which has been proposed as a promising strategy for designing novel functional materials [13].

Typically, a HEA is defined as an alloy containing at least 5 elements with concentrations between 5 and 35 atomic percent, according to the paper by Yeh et al. [14]. Furthermore, Otto et al. suggested that only an alloy that forms a solid solution with no intermetallic phases should be considered as a HEA, because the formation of an ordered phase decreases the entropy of the system [15]. Recently, Koželj et al. discovered superconductivity in $Ta_{34}Nb_{33}Hf_8Zr_{14}Ti_{11}$ with a $T_c$ of 7.3 K [16]. The observed superconductivity in the HEA was conventional phonon-mediated one. Notably, high pressure measurements revealed that the superconductivity (zero resistivity state) in the HEA superconductor was robust at pressure under 190 GPa [17]. This fact suggests that the superconducting states of the HEA can be enhanced by the HEA effect under extreme conditions.

On the basis of the observed structural stability in the HEA superconductor under extremely high pressure, we expected that the local disorder in RE(O,F)BiS$_2$, which negatively affects the emergence of bulk superconductivity, can be suppressed, and the superconducting properties may be enhanced by the HEA effect. However, RE(O,F)BiS$_2$ is a layered system and composed of five different atomic sites, in which a simple HEA concept cannot be applied. Therefore, we focused on the RE site (see the inset of Fig. 1) only and designed the HEA-type RE(O,F) blocking layers by the solid solution of RE = La, Ce, Pr, Nd, and Sm. In this letter, four HEA-type REO$_{0.5}$F$_{0.5}$BiS$_2$ samples were newly synthesized. Powder X-ray diffraction (XRD) and Rietveld refinements revealed that the obtained HEA-type REO$_{0.5}$F$_{0.5}$BiS$_2$ is almost single-phase and structurally homogeneous as in the other typical REO$_{0.5}$F$_{0.5}$BiS$_2$ (RE = La, Ce, Pr, or Nd), where RE site consists of one or two RE elements. For all the HEA-type REO$_{0.5}$F$_{0.5}$BiS$_2$ samples, a sharp superconducting transition was observed. Notably, we observed that the superconducting phase diagram of HEA-type REO$_{0.5}$F$_{0.5}$BiS$_2$ showed a trend clearly different from that of typical REO$_{0.5}$F$_{0.5}$BiS$_2$. Although only the RE site possesses a HEA state in HEA-type REO$_{0.5}$F$_{0.5}$BiS$_2$, the superconducting properties seem to be enhanced as compared to typical REO$_{0.5}$F$_{0.5}$BiS$_2$. Hence, the concept of designing layered superconductors with a HEA-type blocking layer will provide us with a new strategy for synthesizing new layered superconductors with outstanding properties.

We synthesized four REO$_{0.5}$F$_{0.5}$BiS$_2$ superconductors with nominal compositions of La$_{0.3}$Ce$_{0.3}$Pr$_{0.2}$Nd$_{0.1}$Sm$_{0.1}$O$_{0.5}$F$_{0.5}$BiS$_2$, La$_{0.2}$Ce$_{0.2}$Pr$_{0.2}$Nd$_{0.2}$Sm$_{0.2}$O$_{0.5}$F$_{0.5}$BiS$_2$, La$_{0.1}$Ce$_{0.1}$Pr$_{0.3}$Nd$_{0.3}$Sm$_{0.2}$O$_{0.5}$F$_{0.5}$BiS$_2$, and La$_{0.1}$Ce$_{0.1}$Pr$_{0.2}$Nd$_{0.3}$Sm$_{0.3}$O$_{0.5}$F$_{0.5}$BiS$_2$, which are labeled as A, B, C, and D, respectively. The polycrystalline samples were synthesized by the solid-state-reaction method similar to that established for typical REO$_{0.5}$F$_{0.5}$BiS$_2$ [4, 8]. Powders of La$_2$S$_3$ (99.9%), Ce$_2$S$_3$ (99.9%), Pr$_2$S$_3$ (99.9%), Nd$_2$S$_3$ (99%), Sm$_2$S$_3$ (99.9%), Bi$_2$O$_3$ (99.999%), BiF$_3$ (99.9%), grains of Bi (99.999%), and S (99.99%) were used as starting materials. The mixture of the starting materials with a nominal composition listed above was mixed well, pelletized, sealed into an evacuated quartz tube, and heated at 700ºC for 20 h. The obtained sample was ground, mixed, pelletized, sealed into an evacuated quartz tube, and heated at 700ºC for 20 h to homogenize the sample.

The phase purity and the crystal structure were examined using powder X-ray diffraction (XRD) with CuK$\alpha$ radiation by the $\theta$-$2\theta$ method. Rietveld refinements were performed to analyze the obtained XRD data. The RIETAN-FP software was used for the Rietveld analysis [18]. To visualize the refined crystal structure, the VESTA software was used [19]. The actual composition for the RE site was analyzed using energy-dispersive X-ray spectrometry (EDX). As listed in Table S1 (Supplemental data), the analyzed RE concentrations almost corresponded to the nominal values for all the samples. Therefore, in this paper, we describe the sample names using the nominal values for clarity.

The temperature dependence of the electrical resistivity was measured by the four-terminal method with a current of 1 mA. The temperature dependence of the magnetic susceptibility was measured by a superconducting quantum interference device (SQUID) magnetometer after both zero-field cooling (ZFC) and field cooling (FC) with a typical applied filed of 10 Oe.

Figure 1 shows the typical XRD pattern and the Rietveld fitting result for the sample B ($La_{0.2}Ce_{0.2}Pr_{0.2}Nd_{0.2}Sm_{0.2}O_{0.5}F_{0.5}BiS_2$). Although tiny impurity peaks probably due to $RE_2O_2S$ were observed, almost all the peaks were refined using the tetragonal *P*4/*nmm* model [9]. See the Supplementary data for the Rietveld refinement results for A, C, and D. For the sample B, as show in the inset of Fig. 1, the RE site can be regarded as an HEA type with La, Ce, Pr, Nd, Sm with a concentration of 19-21%, which satisfies the definition of the HEA by Yeh et al. [14]. For other samples A, C, and D, high purity of the obtained samples was confirmed by the Rietveld refinements, as well. The obtained crystal structure parameters are listed in Table S1 and S2. Changing the RE concentration from A to D does not affect the structural model with the *P*4/*nmm* space group but systematically decreases the lattice constant of *a* as shown in Fig. 2. This trend is similar to that in the case of $Ce_{1-x}Nd_xO_{0.5}F_{0.5}BiS_2$ or other related systems [18,19]. Generally, in the $REOBiS_2$-type compounds, the lattice expansion/shrinkage is directly linked to the mean ionic radius of the RE site [8,9]. However, the lattice constant of *c* for the HEA samples increases from A to D. On the basis of previous works, we have obtained a trend that the lattice constant of *c* of the $REOBiS_2$-type compounds decreases with increasing carrier concentration [4,8]. Therefore, the increase in *c* for the samples C and D may be related to the decrease in the electron carrier concentration. In the samples C and D, Sm concentration is larger than that in the samples A and B. This may indicate that the Sm ions are in the mixed-valence state of +2 and +3, as observed in other Sm compounds [20]. In addition, the plot of the lattice constant of *a* for the HEA-type samples is clearly larger than that for typical $REO_{0.5}F_{0.5}BiS_2$, which also suggests that the RE ionic radius for any RE site is larger than that of $RE^{3+}$. The in-plane Bi-S1 distance decreases with decreasing *a*. The S1-Bi-S1 angle is almost 180º and does not show a remarkable change between A, B, C, and D. Therefore, with decreasing Bi-S1 distance, in-plane chemical pressure is generated without any change in the flatness of the BiS1 plane [9].

On the homogeneity of the HEA-type samples, there is the possibility of the local phase separation into grains with different RE ions. However, as shown in Fig. 2(c), the typical peak (110 reflection) for the HEA-type samples is not broadened by mixing five RE elements and seems slightly sharper than the typical systems with RE = La and Nd. In addition, the full width half maximum (FWHM) of the 110 peak plotted in Fig. 2(d) shows that the peak sharpness for the HEA-type samples is narrower than that for typical $REO_{0.5}F_{0.5}BiS_2$ (RE = La and Nd). We consider that the RE ions are homogeneously solved in the grains, on the basis of the present powder XRD study. Furthermore, the REO$_{0.5}$F$_{0.5}$BiS$_2$ structure may be stabilized by the HEA effect.

Figure 3(a) shows the temperature dependences of the electrical resistivity for A–D. For all the samples, the resistivity above $T_c$ increases with decreasing temperature, which is a semiconducting-like behavior. In several BiS$_2$-based superconductors, a similar behavior was observed [4,5]. In typical REO$_{0.5}$F$_{0.5}$BiS$_2$, the semiconducting behavior was suppressed with increasing in-plane chemical pressure [10,21]. In addition, in the LaO$_{0.5}$F$_{0.5}$BiS$_{2-x}$Se$_x$, which is the best example to dramatically change the in-plane chemical pressure amplitude, the increase in the in-plane chemical pressure (Se concentration) induced metallic conductivity and bulk superconductivity [10,21]. Although the temperature dependences of the resistivity are quite similar for all the present samples A–D, we notice that the increase in resistivity at low temperature for the C and D samples is weakened than that for A and B in the plot of normalized resistivity [Fig. 3(b)].

Figure 3(c) shows the low-temperature resistivity plots. For all the samples, a sharp superconducting transition was observed. The onset temperature ($T_c^{onset}$) was defined as the temperature where the resistivity begins to decrease upon cooling. The $T_c^{zero}$ was defined as the temperature where the resistivity becomes zero upon cooling. The estimated $T_c^{onset}$ is 3.4, 4.3, 4.7, and 4.9 K for A, B, C, and D, respectively. The estimated $T_c^{zero}$ is 3.0, 3.8, 4.2, and 4.5 K for A, B, C, and D, respectively. The transition width ($\Delta T_c$) estimated from the difference between $T_c^{onset}$ and $T_c^{zero}$ is 0.4, 0.5, 0.5, and 0.4 K, for A, B, C, and D, respectively. The sharpness of the transition for the present sample is sharper than that of typical REO$_{0.5}$F$_{0.5}$BiS$_2$; for example, $\Delta T_c$ of PrO$_{0.5}$F$_{0.5}$BiS$_2$ is ~0.8 K according to the temperature derivative of the resistivity in Ref. 22. In typical REO$_{0.5}$F$_{0.5}$BiS$_2$ systems, the existence of strong superconducting fluctuations has been proposed [23]. Furthermore, the structural instability, due to the presence of lone pair effect of the Bi 6s electrons [11], also exists in the typical REO$_{0.5}$F$_{0.5}$BiS$_2$ system [4,5,24], which is linked to the emergence of bulk superconductivity. These superconducting fluctuations or the structural instabilities can broaden the superconducting transition. On the basis of the discussion above, one can propose that the HEA-type blocking layer may suppress those fluctuations and structural instabilities and induced the sharp superconducting transition.

Figure 3(d) shows the temperature dependences of the magnetic susceptibility $4\pi\chi$ for A, B, C, and D. For all the samples, diamagnetic signals due to the emergence of superconductivity were observed. The transition temperature $T_c^{mag}$ estimated as the temperature where the susceptibility begins to drop is 3.1, 3.8, 4.2, and 4.4 K for A, B, C, and D, respectively. The $T_c^{mag}$ almost corresponds to $T_c^{zero}$ estimated from the resistivity data. For C and D, a large shielding fraction was observed, suggesting the emergence of bulk superconductivity in the HEA-type REO$_{0.5}$F$_{0.5}$BiS$_2$ samples. The magnetization loop for the sample D is shown in Supplementary data. We confirmed that the superconducting states in the HEA-type sample is also robust to the magnetic field. In the typical

REO$_{0.5}$F$_{0.5}$BiS$_2$ systems, the shielding volume fraction is enhanced with increasing transport metallicity [5,10,21]. Therefore, although the suppression is slight as shown in Fig. 3(b), the larger shielding fraction in C and D than that of A and B may be related to the suppression of carrier localization.

Here, we discuss the evolution of $T_c$ and shielding fraction (using $\Delta 4\pi\chi$ calculated by $4\pi\chi$ (FC)-$4\pi\chi$ (ZFC) at 2 K) as a function of the lattice constant of $a$. As mentioned above, the flatness of the BiS1 plane is almost the same for all the samples including typical REO$_{0.5}$F$_{0.5}$BiS$_2$ with RE = La, Ce, Pr, and Nd [plotted in Fig. 3(e,f) for comparison]. Therefore, the lattice constant of $a$ can be a good scale to discuss the shrinkage (in-plane chemical pressure) amplitude of the superconducting BiS1 plane. In Fig. 3(e) and 3(f), the dependences of $T_c$ and $\Delta 4\pi\chi$ as a function of lattice constant of $a$ are plotted with the data of REO$_{0.5}$F$_{0.5}$BiS$_2$ (RE = La, Ce, Pr, and Nd) and the solid-solution systems with RE = La$_{1-x}$Ce$_x$ and Ce$_{1-x}$Nd$_x$ [4,8,12]. The $T_c$ plot for the HEA-type samples (A, B, C, and D) is slightly higher than that for typical REO$_{0.5}$F$_{0.5}$BiS$_2$. The $\Delta 4\pi\chi$ for B, C, and D is close to that for RE = Pr and Nd. However, the $\Delta 4\pi\chi$ for A is notably larger than that expected from the dependence of typical REO$_{0.5}$F$_{0.5}$BiS$_2$ phases. These results may indicate that the HEA-type blocking layer is positively affecting the superconducting states emerging in the BiS$_2$ layers. Probably, in-plane disorder is suppressed by the stabilization of the crystal structure by the HEA effect in REO$_{0.5}$F$_{0.5}$BiS$_2$. To obtain the evidence for the positive link, we will investigate the local crystal structure using synchrotron XRD and X-ray absorption spectroscopy as carried out on the typical BiS$_2$-based superconductor system [9-12].

In conclusion, we newly synthesized four superconducting phases La$_{0.3}$Ce$_{0.3}$Pr$_{0.2}$Nd$_{0.1}$Sm$_{0.1}$O$_{0.5}$F$_{0.5}$BiS$_2$, La$_{0.2}$Ce$_{0.2}$Pr$_{0.2}$Nd$_{0.2}$Sm$_{0.2}$O$_{0.5}$F$_{0.5}$BiS$_2$, La$_{0.1}$Ce$_{0.1}$Pr$_{0.3}$Nd$_{0.3}$Sm$_{0.2}$O$_{0.5}$F$_{0.5}$BiS$_2$, and La$_{0.1}$Ce$_{0.1}$Pr$_{0.2}$Nd$_{0.3}$Sm$_{0.3}$O$_{0.5}$F$_{0.5}$BiS$_2$, whose blocking layer has been designed with the HEA concept. According to the RE concentration and the difference in the RE ionic radius, the lattice constant of $a$ systematically changed in the four samples. For all the samples, a superconducting transition was observed. Notably, the superconducting transition in the resistivity-temperature plot was quite sharp (sharper than that of typical PrO$_{0.5}$F$_{0.5}$BiS$_2$). The sharp transition may indicate that the superconductivity fluctuations and/or the structural instability [4,11,23,24] are suppressed by the HEA effect. The plot of $T_c$ as a function of lattice constant of $a$ for the HEA-type samples was higher than those of typical REO$_{0.5}$F$_{0.5}$BiS$_2$. Furthermore, for the sample A (with a large lattice constant of $a$), the observed shielding volume fraction estimated from the susceptibility data was clearly higher than that expected from the trend in typical REO$_{0.5}$F$_{0.5}$BiS$_2$. The enhancement of the superconducting properties observed in the HEA-type samples may be related to the HEA effect in the blocking layer. We believe that the introduction of the HEA effects into the blocking layers of layered superconductors will be a useful strategy to improve the superconducting properties such as $T_c$, upper critical field, and critical current density.

Acknowledgements

We would like to thank O. Miura for his experimental support. This study was partially supported by the Grants-in-Aid for Scientific Research (Nos. 15H05886, 16H04493, 16K17944, and 17K19058).

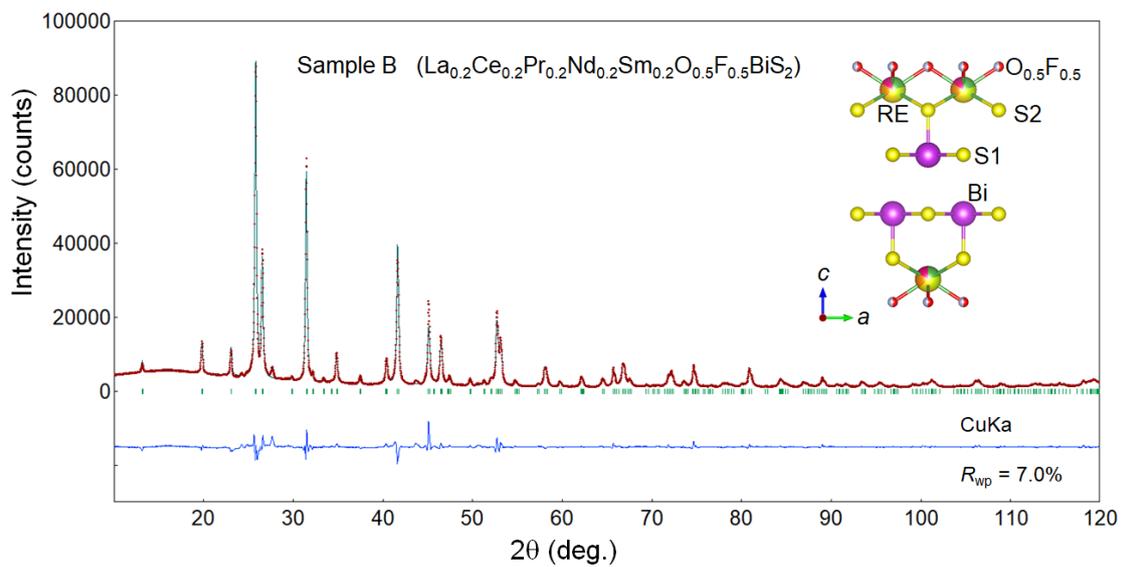

Fig. 1. Powder X-ray diffraction pattern and Rietveld fitting for B (La$_{0.2}$Ce$_{0.2}$Pr$_{0.2}$Nd$_{0.2}$Sm$_{0.2}$O$_{0.5}$F$_{0.5}$BiS$_2$). The inset shows a schematic image of the crystal structure for the sample B.

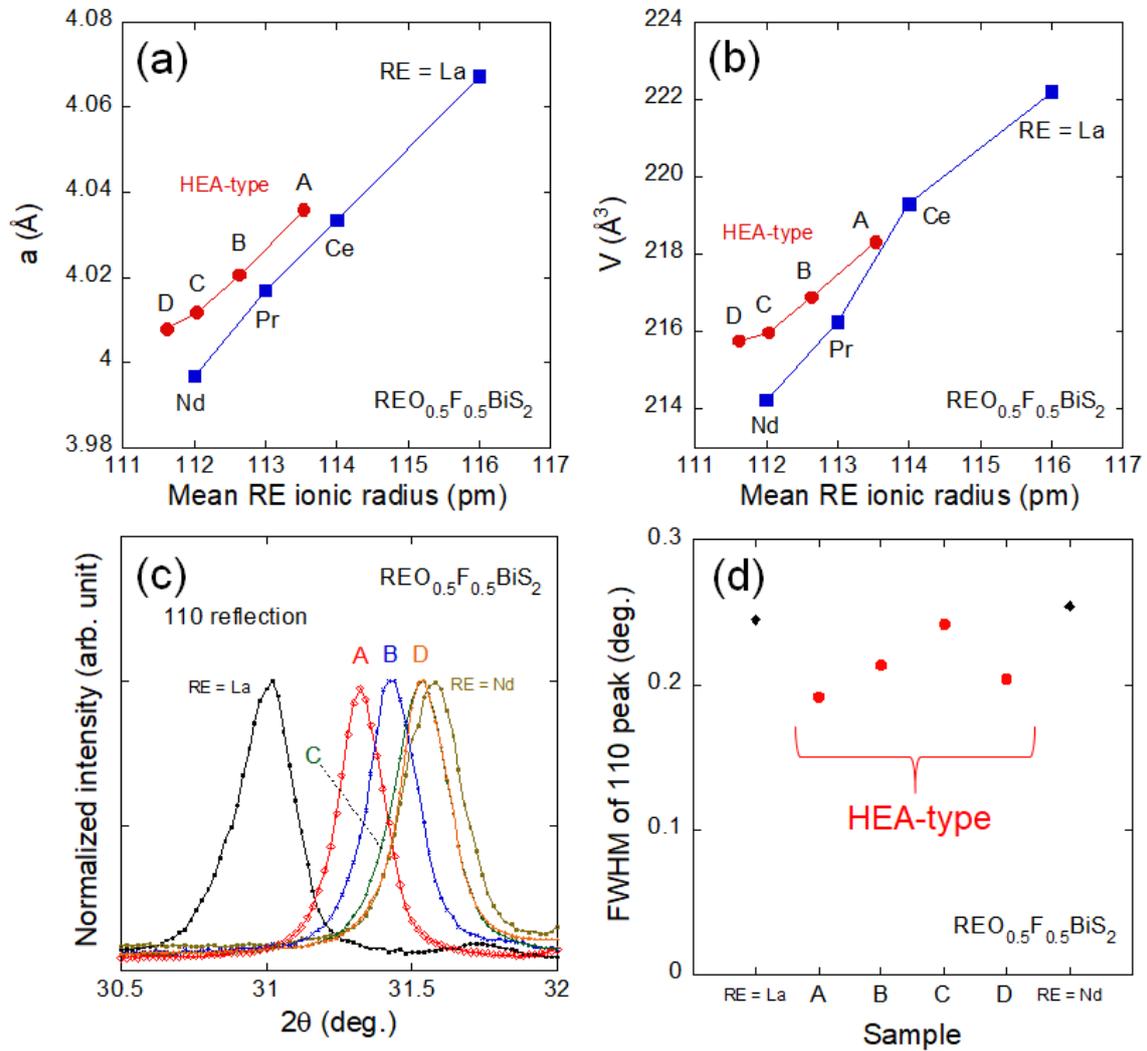

Fig. 2. (a,b) Dependences of lattice constants $a$ and $V$ as a function of mean RE (rare-earth site) ionic radius in typical and high-entropy-ally-type $REO_{0.5}F_{0.5}BiS_2$. For, A, B, C, and D, the analyzed RE concentration was $La_{0.28}Ce_{0.31}Pr_{0.19}Nd_{0.12}Sm_{0.10}$, $La_{0.19}Ce_{0.21}Pr_{0.21}Nd_{0.20}Sm_{0.19}$, $La_{0.09}Ce_{0.09}Pr_{0.30}Nd_{0.32}Sm_{0.20}$, and $La_{0.10}Ce_{0.11}Pr_{0.19}Nd_{0.30}Sm_{0.30}$, respectively. The mean RE ionic radius was calculated using the values for $RE^{+3}$ with a coordination number of 8: 116, 114, 113, and 112 pm for La, Ce, Pr, and Nd. (c) Normalized 110 peaks for A, B, C, D, and RE = La and Nd. (d) Full width half maximum (FWHM) of the 110 peak for A, B, C, D, and RE = La and Nd.

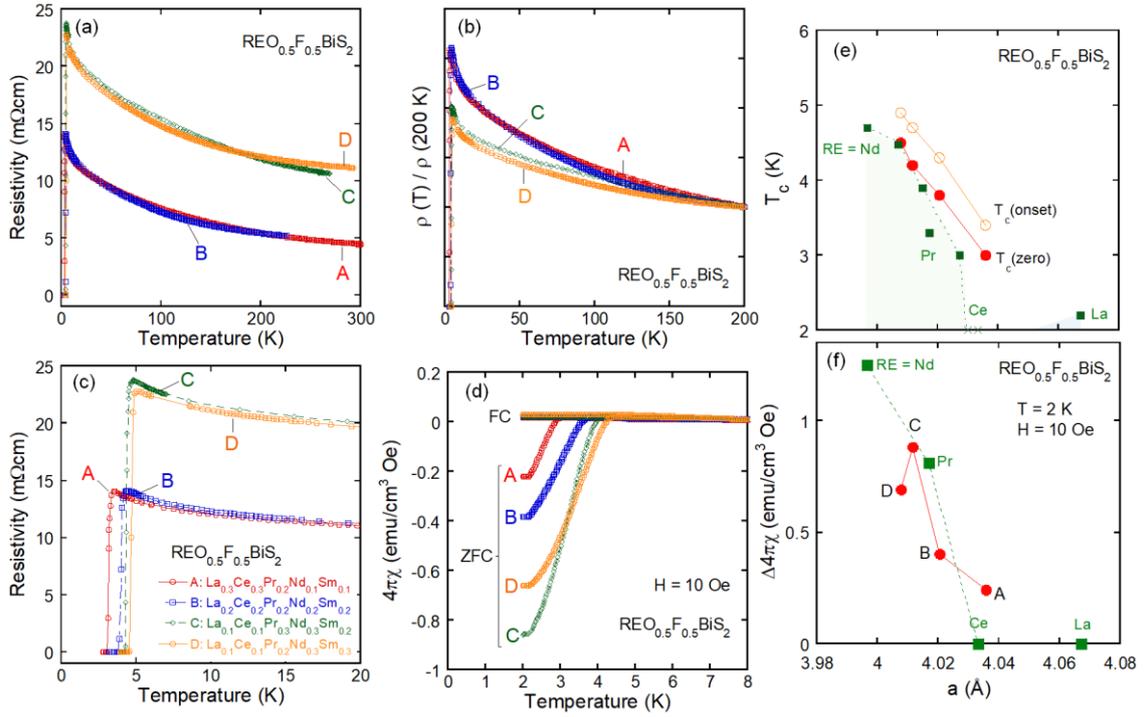

Fig. 3. (a) Temperature dependences of electrical resistivity for A, B, C, and D. (b) Temperature dependences of resistivity normalized at 200 K for A, B, C, and D. (c) Enlarged figure of (a) below 20 K. (d) Temperature dependences of magnetic susceptibility $4\pi\chi$ for A, B, C, and D. (e,f) Dependences of (e) $T_c$ and (f) $\Delta 4\pi\chi$ as a function of lattice constant of $a$. Data for $REO_{0.5}F_{0.5}BiS_2$ (RE = La, Ce, Pr, and Nd) [12] are plotted for comparison. The cross symbols indicate that a superconducting transition is not observed for the samples above 2 K.

# [Supplementary data]

**Superconductivity in REO$_{0.5}$F$_{0.5}$BiS$_2$ with high-entropy-alloy-type blocking layers**

Ryota Sogabe[1], Yosuke Goto[1], Yoshikazu Mizuguchi[1]*

1. Department of Physics, Tokyo Metropolitan University, 1-1, Minami-osawa, Hachioji 192-0397, Japan.

* Corresponding author: Yoshikazu Mizuguchi (mizugu@tmu.ac.jp)

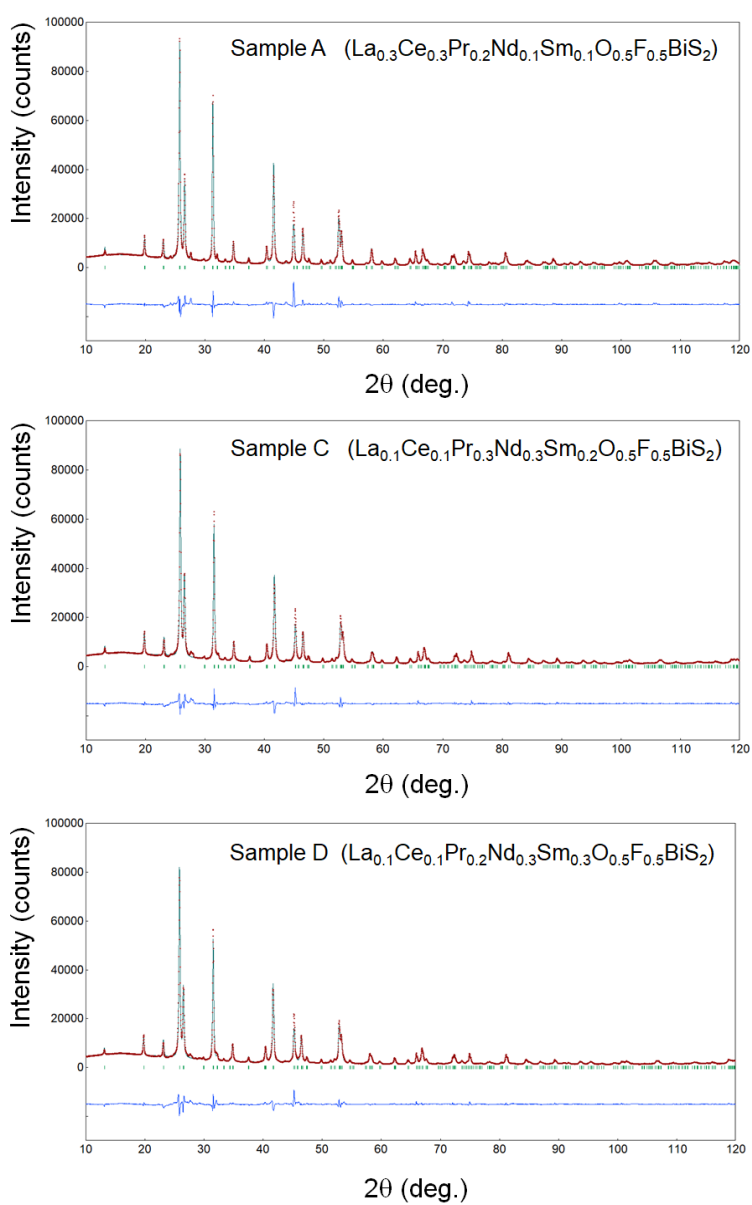

Fig. S1. XRD patterns and Rietveld refinement results for the samples A, C, and D.

**Table. S1. Composition, crystal structure parameter, and superconducting properties of A, B, C, and D.**

| Sample label | A | B | C | D |
|---|---|---|---|---|
| RE (nominal) | $La_{0.3}Ce_{0.3}Pr_{0.2}Nd_{0.1}Sm_{0.1}$ | $La_{0.2}Ce_{0.2}Pr_{0.2}Nd_{0.2}Sm_{0.2}$ | $La_{0.1}Ce_{0.1}Pr_{0.3}Nd_{0.3}Sm_{0.2}$ | $La_{0.1}Ce_{0.1}Pr_{0.2}Nd_{0.3}Sm_{0.3}$ |
| RE (EDX) | $La_{0.28}Ce_{0.31}Pr_{0.19}Nd_{0.12}Sm_{0.10}$ | $La_{0.19}Ce_{0.21}Pr_{0.21}Nd_{0.20}Sm_{0.19}$ | $La_{0.09}Ce_{0.09}Pr_{0.30}Nd_{0.32}Sm_{0.20}$ | $La_{0.10}Ce_{0.11}Pr_{0.19}Nd_{0.30}Sm_{0.30}$ |
| Space group | Tetragonal $P4/nmm$ (No. 129) | | | |
| $a$ (Å) | 4.03587(8) | 4.02046(8) | 4.01173(8) | 4.00785(8) |
| $c$ (Å) | 13.4029(3) | 13.4173(3) | 13.4192(3) | 13.4314(4) |
| $V$ (Å$^3$) | 218.309(8) | 216.879(8) | 215.969(8) | 215.747(9) |
| Bi-S1 distance (Å) | 2.85381(7) | 2.84294(7) | 2.83672(7) | 2.83399(7) |
| S1-Bi-S1 angle (°) | 179.6(3) | 179.4(3) | 179.9(3) | 179.6(4) |
| $R_{wp}$ (%) | 8.1 | 7.0 | 7.1 | 6.9 |
| $T_c^{zero}$ (K) | 3.0 | 3.8 | 4.2 | 4.5 |
| $T_c^{onset}$ (K) | 3.4 | 4.3 | 4.7 | 4.9 |
| $\Delta T_c$ (K) | 0.4 | 0.5 | 0.5 | 0.4 |

**Table S2. Atomic coordinate obtained by Rietveld refinement.** The atomic coordinate is (0, 0.5, z) for RE, Bi, S1 (in-plane S), and S2. The atomic coordinate for the O/F site is (0, 0, 0). We performed Rietveld refinements with fixed displacement parameters, which were obtained from synchrotron XRD study for $PrO_{0.5}F_{0.5}BiS_2$. The occupancy of the RE and O/F sites were fixed to the nominal compositions.

| Sample | A | B | C | D |
|---|---|---|---|---|
| z(RE) | 0.09875(10) | 0.09795(9) | 0.09786(10) | 0.09784(10) |
| z(Bi) | 0.62417(10) | 0.62483(9) | 0.62499(10) | 0.62523(10) |
| z(S1) | 0.3750(5) | 0.3740(5) | 0.3748(5) | 0.3741(6) |
| z(S2) | 0.8144(4) | 0.8156(4) | 0.8149(4) | 0.8146(5) |

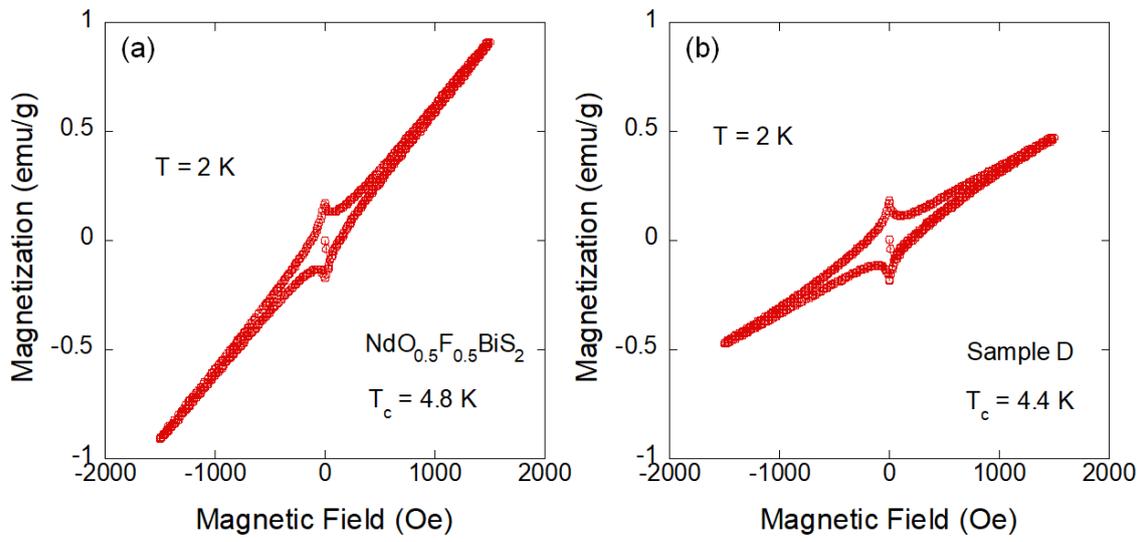

Fig. S2. Magnetic field dependences of the magnetization at 2 K for $NdO_{0.5}F_{0.5}BiS_2$ and the sample D ($La_{0.1}Ce_{0.1}Pr_{0.2}Nd_{0.3}Sm_{0.3}O_{0.5}F_{0.5}BiS_2$).